\documentclass[sigconf,10pt]{acmart}

\pdfoutput=1
\usepackage{booktabs} 
\setcopyright{acmlicensed}
\newcommand{\minihead}[1]{{\vspace{.5em}\noindent\textbf{#1} }}
\newcommand{\red}[1]{{\color{black}#1}}
\newcommand{\mvar}{\red{d}}
\newcommand{\dvar}{\red{n}}

\newcommand\code[1]{\lstinline$#1$}

\renewcommand\paragraph{\@startsection{paragraph}{4}{\z@}%
                                    {0.5ex \@plus 0.5ex \@minus .2ex}%
                                    {-0.5em}%
                                    {\normalfont\normalsize\bfseries}}
\usepackage[labelfont=bf]{caption}
\usepackage{enumitem}

\usepackage{algorithm}
\usepackage[noend]{algpseudocode}
\algdef{SE}[DOWHILE]{Do}{doWhile}{\algorithmicdo}[1]{\algorithmicwhile\ #1}%

\theoremstyle{problem}
\newtheorem{problem}{Problem}[section]

\begin{document}
\title{DROP: A Workload-Aware Optimizer for Dimensionality Reduction}

\author{Sahaana Suri, Peter Bailis}
\affiliation{
  \institution{Stanford University}
}

\renewcommand{\shortauthors}{S. Suri and P. Bailis}

\copyrightyear{2019} 
\acmYear{2019} 
\setcopyright{acmlicensed}
\acmConference[DEEM'19]{International Workshop on Data Management for End-to-End Machine Learning}{June 30, 2019}{Amsterdam, Netherlands}
\acmBooktitle{International Workshop on Data Management for End-to-End Machine Learning (DEEM'30), June 30, 2019, Amsterdam, Netherlands}
\acmPrice{15.00}
\acmDOI{10.1145/3329486.3329490}
\acmISBN{978-1-4503-6797-4/19/06}

\begin{abstract}
Dimensionality reduction (DR) is critical in scaling machine learning pipelines: by reducing input dimensionality in exchange for a preprocessing overhead, DR enables faster end-to-end runtime. Principal component analysis (PCA) is a DR standard, but can be computationally expensive: classically $O(dn^2 + n^3)$ for an $n$-dimensional dataset of $d$ points. 
Theoretical work has optimized PCA via iterative, sample-based stochastic methods. 
However, these methods execute for a fixed number of iterations or to convergence, sampling too many or too few datapoints for end-to-end runtime improvements. 
We show how accounting for downstream analytics operations during DR via PCA allows stochastic methods to efficiently terminate after processing small (e.g., 1\%) samples of data. 
Leveraging this, we propose DROP, a DR optimizer that enables speedups of up to \red{$5\times$} over \red{Singular-Value-Decomposition (SVD)-based} PCA, and \red{$16\times$} over conventional DR methods in end-to-end nearest neighbor workloads.
\end{abstract}

\maketitle

\section{Introduction}
\label{sec:intro}

Rapid growth in high-dimensional data from automated sources poses a scalability challenge for machine learning pipelines~\cite{plato,macrobase-cidr}.
Practitioners turn to dimensionality reduction (DR) techniques to alleviate this challenge~\cite{keogh-indexing,local-dr,decade,gemini}.
DR methods transform an $n$-dimensional dataset to a lower $k$-dimensional representation while preserving salient dataset features.
This allows downstream analytics routines to run in time that scales with $k$ while preserving downstream task accuracy.
Thus, in exchange for a preprocessing overhead, DR techniques can decrease end-to-end workload runtime.


To attain a target downstream task accuracy with as low a transformation dimension $k$ as possible, Principal Component Analysis (PCA) is often practitioners' DR method of choice ~\cite{jolbook}. 
However, classic task-independent PCA implementations (i.e., via Singular Value Decomposition or SVD) scale poorly, with overheads that outweigh DR's downstream runtime benefit. 
In response, practitioners may use faster DR techniques that return higher $k$ for a given accuracy, but provide a lower end-to-end runtime. 
For instance, PCA is traditionally considered too computationally expensive for time series similarity search~\cite{time-series-dm}, with widely-cited work excluding it for more efficient,  less precise methods~\cite{keogh-study}. 

In this work, we accelerate PCA by leveraging the fact that practitioners are willing to trade dimensionality for whole workload runtime. 
We build off the key insight from theoretical means of optimizing PCA via stochastic, or sample-based methods~\cite{shamir,re-new}:  instead of processing the entire dataset at once, data samples can be used to iteratively refine an estimate of the transformation until the transformation converges. 
However, for end-to-end runtime optimization, sampling to convergence can be unnecessary as DR for downstream tasks such as similarity search are effective even with suboptimal, higher-dimensional transformations~\cite{keogh-study}.
Taking into account this workload-dependent tolerance to error during sample-based PCA would enable us to terminate prior to convergence, and thus more quickly return an accuracy-preserving transformation. 
This leaves the open question of: given a dataset and workload, what sampling rate (i.e., termination condition) is required to optimize for both accuracy and end-to-end runtime?

To resolve this data- and workload-dependent trade-off, we develop DROP, a system that dynamically identifies the amount of sampling required for stochastic PCA by using downstream task information.
DROP takes as input a high-dimensional dataset,
a workload-dependent constraint on approximation accuracy (e.g., pairwise Euclidean distance to 5\% for similarity search, see \S~\ref{sec:RQW}), and an optional runtime model expressing downstream computational cost as a function of dimensionality (e.g., for k-Nearest Neighbors [k-NN], runtime is linear in dimensionality). 
DROP returns a low-dimensional transformation for the input using as few samples as needed to minimize the projected overall workload runtime while preserving the input constraint.

DROP addresses the question of how much to sample the input dataset via data-dependent progressive sampling and online progress estimation at runtime.  
DROP performs PCA on a small sample to obtain a candidate transformation, then increases the number of samples until termination. 
To identify the termination point that minimizes runtime, DROP must overcome three challenges:

First, given the results of PCA on a data sample, DROP must \emph{evaluate the quality} of the current candidate transformation.
Popular analytics and data mining tasks often require approximate preservation of metrics such as average pairwise distances between data points~\cite{time-series-dm,dm-book}, which are costly to compute.
Thus, DROP adapts confidence intervals for fast estimation of the input metric to preserve.

Second, DROP must \emph{estimate the marginal benefit of sampling additional datapoints}.
When running PCA on a series of larger samples, later samples will increase DR runtime, but may return lower $k$ (lower downstream runtime)---DROP must estimate how these values will change in future iterations to navigate this trade-off between end-to-end runtime and dimensionality.
DROP uses the results obtained from previous iterations to fit predictive models for dimensionality and runtime of the next iteration.

Finally, given the predicted marginal benefit, DROP must \emph{optimize end-to-end runtime}.
While an application-agnostic approach would iterate until successive iterations yield little or no benefit, a user-provided runtime model may reveal that trading a higher $k$ for a lower DR runtime may decrease overall runtime.
DROP evaluates the runtime model at each iteration
to minimize the expected workload runtime.

DROP is a system that combines recent theoretical advances in DR and classic techniques from approximate query processing for end-to-end workflow optimization.
In this work, we make the following contributions:
\begin{itemize}

\item We show the data sample required to perform accuracy-achieving PCA is often small (as little as $1\%$), and sampling can enable up to \red{$91\times$} speedup over baseline PCA. 
  
\item We propose DROP, an online optimizer for DR that uses information about downstream analytics tasks to perform efficient stochastic PCA.

\item We present techniques based on progressive sampling, approximate query processing, online progress estimation, and cost based optimization to enable up to \red{$5\times$} faster end-to-end execution over PCA via SVD.
\end{itemize}

\section{Related Work}
\label{sec:relwork}
\label{sec:relatedwork}

\minihead{Dimensionality Reduction} DR is a well-studied operation~\cite{dr-survey1,dr-survey2,nonlinear-dr} in the
database~\cite{keogh-indexing,local-dr,charu-ss}, data
mining~\cite{sax,paa}, statistics and machine
learning~\cite{alecton,shamir,bernstein} communities.
In this paper, our focus is on DR via PCA.
While classic PCA via SVD is inefficient, stochastic~\cite{re-new, shamir} and randomized~\cite{tropp} methods provide scalable alternatives.
DROP draws from both the former to tackle the challenge of how much data to sample, and the latter for its default PCA operator (though DROP's modular architecture makes it simple to use any method in its place). Further, to the best of our knowledge, these advanced methods for PCA have not been empirically compared head-to-head with conventional DR approaches such as Piecewise Approximate Averaging~\cite{paa}.

\minihead{Approximate Query Processing (AQP)} 
Inspired by AQP engines~\cite{barzan-keynote}
as in online aggregation~\cite{onlineagg}, DROP performs progressive
sampling.  
While DROP performs simple uniform sampling, the literature contains a wealth of techniques for various biased sampling techniques~\cite{surajit-sample, surajit-2}.
DROP performs online progress estimation to minimize the
end-to-end analytics cost function. This is analogous to query
progress estimation~\cite{qpi1} and performance
prediction~\cite{mr-predict} in database and data
warehouse settings and has been exploited in approximate query
processing engines such as BlinkDB~\cite{blinkdb}. 

\minihead{Scalable \red{ Workload-Aware, }Complex Analytics} DROP is an operator
for analytics dataflow pipelines. Thus, DROP is
an extension of recent results on integrating complex
analytics function including model training~\cite{bismarck,mcdb} and
data exploration~\cite{scorpion,canopy,kraska-viz} operators into analytics engines. 

\begin{figure*}
\begin{center}
\includegraphics[width=\textwidth]{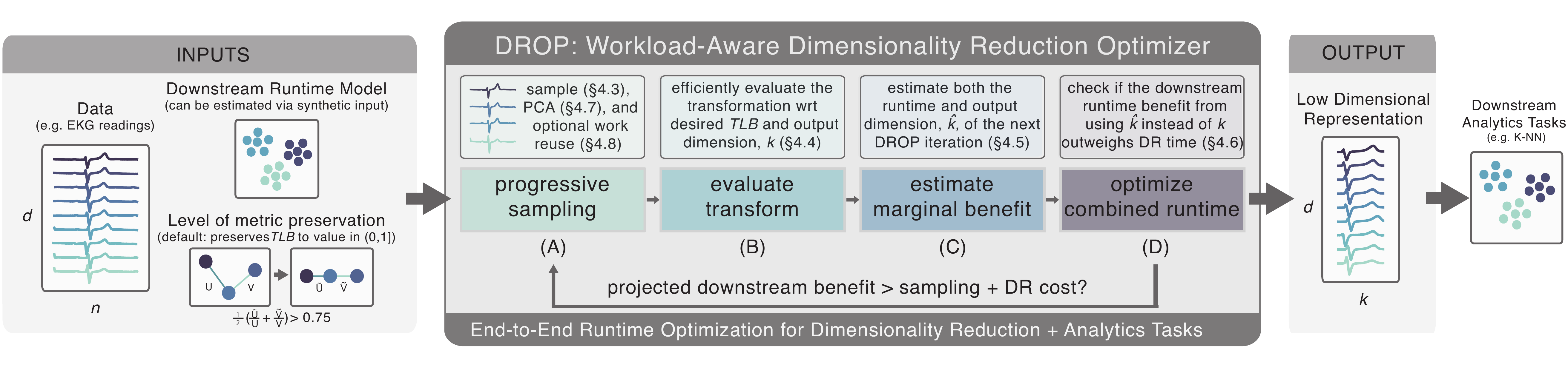}\vspace{-1em}
\caption[]{High-level DROP architecture depicting DROP's inputs, outputs, and core components.}
\end{center}
\vspace{-1em}
\label{fig:arch}
\end{figure*}

\section{Background and Problem}
\label{sec:background}

In this section, we provide background on dimensionality reduction (DR) and our problem of workload-aware DR.

\subsection{Dimensionality Reduction}
\label{sec:defs}

The goal of DR is to find a low-dimensional representation of a dataset that preserves metrics of interest, such as data point similarity~\cite{dr-survey1,dr-survey2}. Formally, consider a data matrix $X \in \mathbb{R}^{\mvar \times \dvar}$, where each row $i$ corresponds to data point $x_i \in \mathbb{R}^\dvar$, with $\mvar > \dvar$.  
DR computes a transformation function $T: \mathbb{R}^\dvar \rightarrow \mathbb{R}^k$ that maps each $x_i$ to a more compact representation, resulting in a new data matrix $T(X) = \tilde{X} \in \mathbb{R}^{\mvar \times k}$.

\subsubsection*{Principal Component Analysis (PCA)}
\label{sec:pca}
PCA is a linear DR technique that identifies a new orthogonal basis for a dataset that captures its directions of highest variance.
Of all linear transformations, this basis minimizes reconstruction error in a mean square sense. 
Classically implemented PCA uses a Singular Value Decomposition (SVD) routine~\cite{trefethen}.

\subsection{DR \red{for Repeated-Query Workloads}}
\label{sec:RQW}

In workloads such as similarity search, clustering, \red{or classification}, ML models are periodically trained over historical data, and are \emph{repeatedly queried} as incoming data arrives or new query needs arise. 
Indexes built over this data can improve the efficiency of this repeated query workload in exchange for a preprocessing overhead.
DR with a multidimensional index structure in the reduced space is a classic way of achieving this~\cite{local-dr,dynamic-ss,dm-book,decade,search}.

\subsubsection*{\red{DR in Similarity Search}}
Similarity search is a repeated-query workload performed over data types including images, documents and time series~\cite{keogh-study,lsh}.
When similarity is measured by Euclidean distance, our goal is to find a low-dimensional representation that approximately preserves pairwise $\ell_2$-distances between data points. We quantify this distance-preservation property using the Tightness of Lower Bounds ($TLB$) metric~\cite{keogh-study}, which estimates the performance difference in a downstream similarity search routine (i.e., k-Nearest Neighbors) after DR \emph{without} running the routine:  
\begin{equation}
\label{eq:tlb}
TLB = \frac{2}{\mvar(\mvar-1)}\sum_{i<j}\frac{\| \tilde{x}_i -  \tilde{x}_j \|_2 }{\| x_i -  x_j\|_2 }.
\end{equation}

Given the large amount of research in the space, we use time series similarity search as a running case study throughout this paper. 
We briefly revisit a comparison of DR techniques for time series similarity search from VLDB 2008~\cite{keogh-study} to verify that PCA can outperform conventionally used techniques (low $k$), but with a high DR runtime cost.
The authors omit PCA due to it being ``untenable for large data sets." 

We compare PCA via SVD to baseline techniques based on returned dimensionality and runtime with respect to $TLB$ over the largest datasets from~\cite{keogh-study}. 
We use their two fastest methods as our baselines as they show the remainder exhibited ``very little difference'': Fast Fourier Transform (FFT) and Piecewise Aggregate Approximation (PAA).
On average, PCA admits an output dimension $k$ that is $2.3\times$ (up to $3.9\times$) and $3.7\times$ (up to $26\times$) smaller than PAA and FFT for $TLB = 0.75$, and $2.9\times$ (up to $8.3\times$) and $1.8\times$ (up to $5.1\times$) smaller for $TLB = 0.99$.
However, PCA implemented via out-of-the-box SVD is on average over \red{$26\times$ (up to $56\times$)} slower than PAA and over \red{$4.6\times$ (up to $9.7\times$)} times slower than FFT when computing the smallest $TLB$-preserving basis.
While the margin between PCA and alternatives is dataset-dependent, PCA almost always preserves $TLB$ with a lower dimensional representation at a higher runtime cost.
This runtime-dimensionality trade-off motivates our study of workload-aware DR methods.

\subsection{Problem: Workload-Aware DR}
\label{subsec:wadr}

In workload-aware DR, we perform DR to minimize workload runtime subject to downstream metric constraints.
DR is a fixed cost (i.e., index construction for similarity search), while workload queries incur a marginal cost dependent on DR dimensionality (i.e., nearest neighbor query). 

As input, consider a dataset $X$, target metric preservation or constraint on approximation accuracy $B$ (e.g., $TLB \geq .99$), and optional downstream runtime model as a function of dimensionality $\mathcal{C}_\mvar(\dvar)$ for an $\mvar\times \dvar$ matrix.  
The metric provides insight into how DR affects downstream accuracy by characterizing workload accuracy without requiring the workload to be run (i.e., $TLB$ for similarity search). 
Denoting DR runtime as $R$, we define the problem:
\begin{problem}
\label{def:opt}
  Given $X \in \mathbb{R}^{\mvar \times \dvar}$, $TLB$ constraint $B \in 
  (0, 1]$, confidence $c$, and workload runtime function $\mathcal{C}_\mvar:\mathbb{Z}_{+} \rightarrow \mathbb{R}_{+}$, find $k$ and transformation
  matrix $T_k \in \mathbb{R}^{\dvar \times k}$ that minimizes $R + \mathcal{C}_\mvar(k)$
  such that $TLB(XT_k) \geq B$ with confidence $c$.
\end{problem}

We assume $\mathcal{C}_\mvar(\dvar)$ is monotonically increasing in $\dvar$.
The more DR time spent, the smaller the transformation (as in the case study), thus the lower the workload runtime.
To minimize $R + \mathcal{C}_\mvar(k)$, we  determine how much time to spend on DR (thus, what $k$ to return) to minimize overall runtime.

\section{DROP: Workload Optimization}
\label{sec:algo}

In this section, we introduce DROP, a system that performs workload-aware DR via progressive sampling and online progress estimation.
DROP takes as input a target dataset, metric to preserve (default, target $TLB$), and an optional downstream runtime model.
DROP then uses sample-based PCA to identify and return a low-dimensional representation of the input that preserves the specified property while minimizing estimated workload runtime (Figure 2, Alg.~\ref{alg:DROP}).

\begin{figure}
\includegraphics[width=\linewidth]{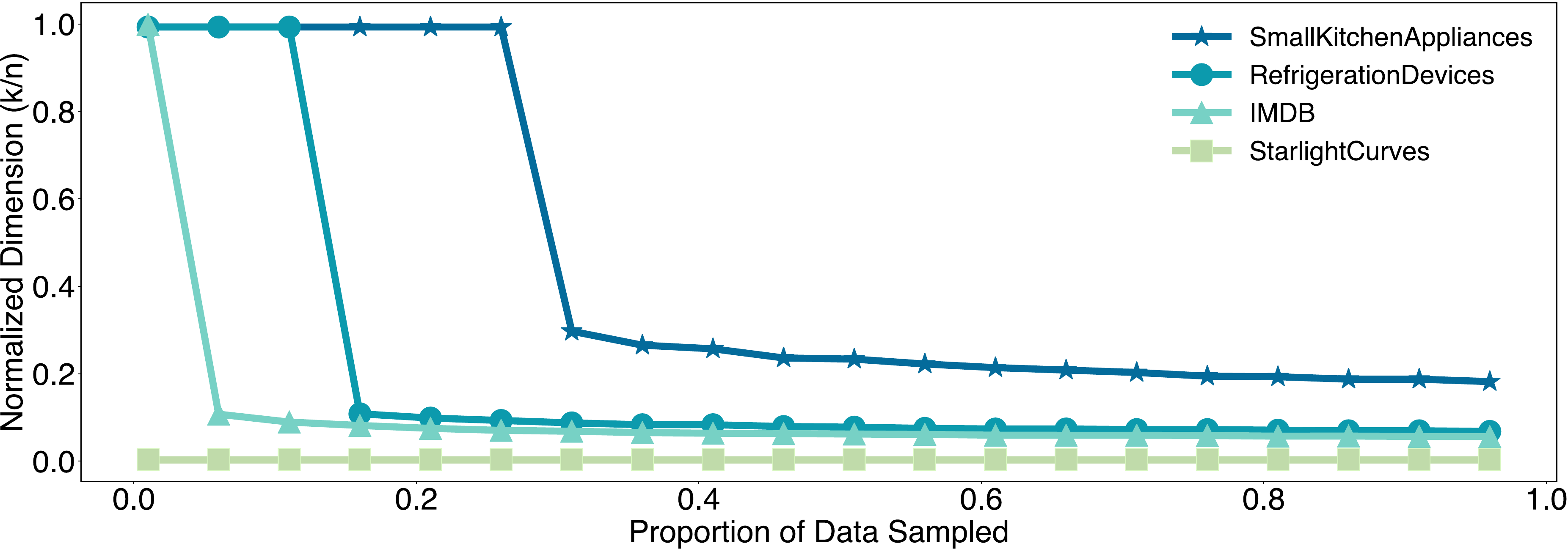}
\caption[]{ Reduction in dimensionality for  $TLB = 0.80$ with progressive sampling. Dimensionality decreases until reaching a state equivalent to running PCA over the full dataset ("convergence").}
\label{fig:progressive}
\end{figure}

\subsection{DROP Algorithm}
\label{subsec:arch}
DROP operates over a series of data samples, and determines when to terminate via \red{a} four-step procedure at each iteration: 

\minihead{Step 1: Progressive Sampling (\S\ref{subsec:psample})}

\noindent DROP draws a data sample, performs PCA over it, and uses a novel reuse mechanism across iterations (\S\ref{subsec:reuse}).

\minihead{Step 2: Transform Evaluation (\S\ref{subsec:teval})} 

\noindent DROP evaluates the above by identifying the size of the smallest metric-preserving transformation that can be extracted. 

\minihead{Step 3: Progress Estimation (\S\ref{subsec:pest})} 

\noindent Given the size of the smallest metric-preserving transform and the time required to obtain this transform, DROP estimates the size and computation time of continued iteration.

\minihead{Step 4: Cost-Based Optimization (\S\ref{subsec:opt})} 

\noindent DROP optimizes over DR and downstream task runtime to determine if it should terminate.

\subsection{Progressive Sampling}
\label{subsec:psample}

Inspired by stochastic PCA methods (\S\ref{sec:relatedwork}), DROP uses sampling to tackle workload-aware DR. 
Many real-world \red{datasets} are intrinsically low-dimensional; a small data sample is sufficient to characterize dataset behavior. 
To verify, we extend our case study (\S\ref{sec:RQW}) by computing how many uniformly selected data samples are required to obtain a $TLB$-preserving transform with $k$ equal to input dimension $\dvar$.
On average, a sample of under $0.64\%$ $(\text{up to } 5.5\%)$ of the input is sufficient for $TLB = 0.75$, and under $4.2\%$ $(\text{up to } 38.6\%)$ is sufficient for $TLB=0.99$.  
If this sample rate is known a priori, we obtain up to \red{$91\times$ speedup} over PCA via SVD.

However, this benefit is dataset-dependent, and unknown a priori.
We thus turn to progressive sampling (gradually increasing the sample size) to identify how large a sample suffices.
Figure~\ref{fig:progressive} shows how the dimensionality required to attain a given $TLB$ changes when we vary dataset and proportion of data sampled.
Increasing the number of samples (which increases PCA runtime) provides lower $k$ for the same $TLB$.
However, this decrease in dimension plateaus as the number of samples increases.
Thus, while progressive sampling allows DROP to tune the amount of time spent on DR, DROP must determine when the downstream value of decreased dimension is overpowered by the cost of DR---that is, whether to sample to convergence or terminate early (e.g., at $0.3$ proportion of data sampled for SmallKitchenAppliances).

Concretely, DROP first repeatedly chooses a subset of data and computes a $\dvar$-dimensional transformation via PCA on the subsample, and then proceeds to determine if continued sampling is beneficial to end-to-end runtime.
We consider a simple uniform sampling strategy: each iteration, DROP samples a fixed percentage of the data.



\begin{algorithm}[t!]
\begin{algorithmic}[1]
\small
\Statex \textbf{Input:}  $X$: data; $B$: target metric preservation level; $\mathcal{C}_\mvar$: cost of downstream operations
\Statex \textbf{Output:} $T_k$: $k$-dimensional transformation matrix
\Statex
\Statex \hrule
\Function{drop}{$X,  B, \mathcal{C}_\mvar$}:
	\State Initialize: $i = 0; k_0 = \infty$ 
		\Comment{iteration and current basis size}
	\Do
		\State i$\texttt{++}$, \textsc{clock.restart}
		\State $X_i$ = \textsc{sample}($X, \textsc{sample-schedule}(i)$) \label{eq:sample}
			\Comment{\S~\ref{subsec:psample}}
		\State $T_{k_i}$ = \textsc{compute-transform}($X, X_i,  B$) \label{eq:evaluate}
			\Comment{\S~\ref{subsec:teval}}
		\State $r_i = \textsc{clock.elapsed}$	
			\Comment{$R = \sum_i r_i$}
		\State $\hat{k}_{i+1}, \hat{r}_{i+1} $ = \textsc{estimate}($k_i, r_i$) \label{eq:estimate}
			\Comment{\S~\ref{subsec:pest}}
	\doWhile{\textsc{optimize}($\mathcal{C}_\mvar,k_i,r_i,\hat{k}_{i+1}, \hat{r}_{i+1}$)} \label{eq:optimize}
		\Comment{\S~\ref{subsec:opt}}
	\\\Return{$T_{k_i}$}
\EndFunction
\end{algorithmic}
\caption{DROP Algorithm}
\label{alg:DROP}
\end{algorithm}

\subsection{Transform Evaluation}
\label{subsec:teval}
DROP must accurately and efficiently evaluate this iteration's performance with respect to the metric of interest \red{over the entire dataset}. 
We define this iteration's performance as the size of the lowest dimensional $TLB$-preserving transform ($k_i$) that it can return. 
There are two challenges in performance evaluation.
First, the lowest $TLB$-achieving $k_i$ is unknown a priori. 
Second, brute-force $TLB$ computation would dominate the runtime of computing PCA over a sample. 
We now describe how to solve these challenges.

\subsubsection{Computing the Lowest Dimensional Transformation}

Given the $\dvar$-dimensional transformation from step 1, to reduce dimensionality, DROP must determine if a smaller dimensional $TLB$-preserving transformation can be obtained and return the smallest such transform. 
Ideally, the smallest $k_i$ would be known a priori, but in practice, this is not true---thus, DROP uses the $TLB$ constraint and two properties of PCA to automatically identify it.

First, PCA via SVD produces an orthogonal linear transformation where the principal components  are returned in order of decreasing dataset variance explained.
As a result, once DROP has computed the transformation matrix for dimension $\dvar$, DROP obtains the transformations for all dimensions $k$ less than $\dvar$ by truncating the matrix to $\dvar \times k$ .

Second, with respect to $TLB$ preservation, the more principal components that are retained, the better the lower-dimensional representation in terms of $TLB$.  
This is because orthogonal transformations such as PCA preserve inner products. 
Therefore, an $\dvar$-dimensional PCA perfectly preserves $\ell_2$-distance between data points. 
As $\ell_2$-distance is a sum of squared (positive) terms, the more principal components retained, the better the representation preserves $\ell_2$-distance.

Using the first property, DROP obtains all low-dimensional transformations for the sample from the $\dvar$-dimensional basis.  
Using the second property, DROP runs binary search over these transformations to return the lowest-dimensional basis that attains $B$ (Alg.~\ref{alg:candidate}, l\ref{eq:basis}).
If $B$ cannot be realized with this sample, DROP omits further optimization steps and continues the next iteration by drawing a larger sample.

Additionally, computing the full $\dvar$-dimensional basis at every iteration may be wasteful. 
Thus, if DROP has found a candidate $TLB$-preserving basis of size $\dvar' < \dvar$ in prior iterations, then DROP only computes $\dvar'$ components at the start of the next iteration.
This allows for more efficient PCA computation for future iterations, as advanced PCA routines can exploit the $\dvar'$-th eigengap to converge faster (\S\ref{sec:relatedwork}).


\subsubsection{Efficient $TLB$ Computation}

Given a transformation, DROP must determine if it preserves the desired $TLB$.
Computing pairwise $TLB$ for all data points requires $O(\mvar^2\dvar)$ time, which dominates the runtime of computing PCA on a sample.
However, as the $TLB$ is an average of random variables bounded from 0 to 1, DROP can use sampling and confidence intervals to compute the $TLB$ to arbitrary confidences.

Given a transformation, DROP iteratively refines an estimate of its $TLB$ (Alg.~\ref{alg:candidate}, l\ref{eq:eval}) by \red{incrementally sampling an increasing number of} pairs from the input data (Alg.~\ref{alg:candidate}, l\ref{eq:paircheck}), transforming each pair into the new basis, then measuring the distortion of $\ell_2$-distance between the pairs, providing a $TLB$ estimate to confidence level $c$ (Alg.~\ref{alg:candidate}, l\ref{eq:tlbeval}). 
If the confidence interval's lower bound is greater than the target $TLB$, the basis is a sufficiently good fit; if its upper bound is less than the target $TLB$, the basis is not a sufficiently good fit. 
If the confidence interval contains the target $TLB$,  \red{ DROP cannot determine if the target $TLB$ is achieved. 
Thus, DROP automatically samples additional pairs to refine its estimate.
}

To estimate the $TLB$ to confidence $c$, DROP uses the Central Limit Theorem: computing the standard deviation of a set of sampled pairs' $TLB$ measures and applying a confidence interval to the sample according to the $c$.

The techniques in this section are presented in the context of $TLB$, but can be applied to any downstream task and metric for which we can compute confidence intervals and are monotonic in number of principal components retained.

\begin{algorithm}
\begin{algorithmic}[1]
\small
\Statex \textbf{Input:}  
\Statex $X$: sampled data matrix
\Statex $B$: target metric preservation level; default $TLB = 0.98$
\Statex  \hrule 
\Function{compute-transform}{$X, X_i B$}: \label{eq:basis}
	\State \textsc{pca.fit}$(X_i)$
			\Comment{fit PCA on the sample}
	\State Initialize: high $= k_{i-1}$; low $=0$; $k_i= \frac{1}{2}$(low + high); $B_i = 0$
	\While{(low $!=$ high)}
		\State $T_{k_i}, B_i  = \textsc{evaluate-tlb}( X, B, k_i)$
		\If{$B_i \leq B$}  low $= k_i + 1$ 
		\Else  \hspace{0pt} high $= k_i $
		\EndIf
		\State $k_i = \frac{1}{2}$(low + high)
	\EndWhile
	\State $T_{k_i} = $ cached $k_i$-dimensional PCA transform\\
	\Return $T_{k_i}$
\EndFunction
\Statex 
\Function{evaluate-tlb}{$X, B, k$}: \label{eq:eval}
	\State numPairs $= \frac{1}{2}\mvar(\mvar-1)$
	\State $p = 100$
		\Comment{number of pairs to check metric preservation}
	\While{($p < $ numPairs)}
		\State $B_i, B_{lo}, B_{hi} = $ \textsc{tlb}($ X, p, k$)
			 \label{eq:paircheck}
		\If{($B_{lo} > B$ or $B_{hi} < B$)}   \textbf{break}
		\Else \hspace{0pt} pairs $\times$= $ 2$
		\EndIf
	\EndWhile
	\\\Return $B_i$	
\EndFunction
\Statex 
\Function{tlb}{$X, p, k$}: \label{eq:tlbeval}
	\State \textbf{return } mean and 95\%-CI of the $TLB$ after transforming $p$ $d$-dimensional pairs of points from $X$ to dimension $k$. The highest transformation computed thus far is cached to avoid recomputation of the transformation matrix.
\EndFunction

\end{algorithmic}
\caption{Basis Evaluation and Search}
\label{alg:candidate}
\end{algorithm}

\subsection{Progress Estimation}
\label{subsec:pest}

Recall that the goal of workload-aware DR is to minimize $R + \mathcal{C}_\mvar(k)$ such that $TLB(XT_k) \geq B$, with $R$ denoting total DR (i.e., DROP's) runtime, $T_k$ the $k$-dimensional $TLB$-preserving transformation of data $X$ returned by DROP, and $\mathcal{C}_\mvar(k)$ the workload cost function. 
Therefore, given a $k_i$-dimensional transformation $T_{k_i}$ returned by the evaluation step of DROP's $i^{\text{th}}$ iteration, DROP can compute the value of this objective function by substituting its elapsed runtime for $R$ and $T_{k_i}$ for $T_k$.  
We denote the value of the objective at the end of iteration $i$ as $obj_i$. 

To decide whether to continue iterating to find a lower dimensional transform, we show in  \S\ref{subsec:opt} that DROP must estimate $obj_{i+1}$. To do so, DROP must estimate the runtime required for iteration $i+1$ (which we denote as $r_{i+1}$, where $R=\sum_i r_i$ after $i$ iterations) and the dimensionality of the $TLB$-preserving transformation produced by iteration $i+1$, $k_{i+1}$. 
DROP cannot directly measure $r_{i+1}$ or $k_{i+1}$ without performing iteration $i+1$, thus performs online progress estimation. Specifically, DROP performs online parametric fitting to compute future values based on prior values for $r_{i}$ and $k_i$ (Alg.~\ref{alg:DROP}, l\ref{eq:estimate}). 
By default, given a sample of size $m_i$ in iteration $i$, DROP performs linear extrapolation to estimate $k_{i+1}$ and $r_{i+1}$. The estimate of $r_{i+1}$, for instance, is:

\vspace{-.4cm}
\begin{equation*}
\hat{r}_{i+1} = r_i + \frac{r_i - r_{i-1}}{m_i - m_{i-1}} (m_{i+1} -  m_i).
\end{equation*}

\subsection{Cost-Based Optimization}
\label{subsec:opt}

DROP must determine if continued PCA on additional samples will improve overall runtime. 
Given predictions of the next iteration's runtime ($\hat{r}_{i+1}$) and dimensionality ($\hat{k}_{i+1}$), DROP uses a greedy heuristic to estimate the optimal stopping point.
If the estimated objective value is greater than its current value ($obj_i < \widehat{obj}_{i+1}$), DROP will terminate. 
If DROP's runtime is convex in the number of iterations, we can prove that this condition is the optimal stopping criterion via convexity of composition of convex functions. 
This stopping criterion leads to the following check at each iteration (Alg.\ref{alg:DROP}, l\ref{eq:optimize}): 

\vspace{-.4cm}
\begin{align}
  obj_i &< \widehat{obj}_{i+1} \nonumber \\
  \mathcal{C}_\mvar(k_i) + \sum_{j=0}^i r_j &< \mathcal{C}_\mvar(\hat{k}_{i+1}) + \sum_{j=0}^{i} r_j + \hat{r}_{i+1} \nonumber \\
  \mathcal{C}_\mvar(k_i) - \mathcal{C}_\mvar(\hat{k}_{i+1}) &< \hat{r}_{i+1}  \label{eq:check}
\end{align}

DROP terminates when the projected time of the next iteration exceeds the estimated downstream runtime benefit. 

\subsection{Choice of PCA Subroutine}
\label{subsec:pcaroutine}

The most straightforward means of implementing PCA via SVD in DROP is computationally inefficient compared to DR alternatives (\S\ref{sec:background}).  
DROP computes PCA via a randomized SVD algorithm from~\cite{tropp} (SVD-Halko).
Alternative efficient methods for PCA exist (i.e., PPCA, which we also provide), but we found that SVD-Halko is asymptotically of the same running time as techniques used in practice, is straightforward to implement, is $2.5-28\times$ faster than our baseline implementations of SVD-based PCA, PPCA, and Oja's method, and does not require hyperparameter tuning for batch size, learning rate, or convergence criteria.  

\subsection{Work Reuse}
\label{subsec:reuse}

A natural question arises due to DROP's iterative architecture: can we combine information across each sample's transformations without computing PCA over the union of the data samples? 
Stochastic PCA methods enable work reuse across samples as they iteratively refine a single transformation matrix, but other methods do not.
DROP uses two insights to enable work reuse over any PCA routine.

First, given PCA transformation matrices $T_1$ and $T_2$, their horizontal concatenation $H = [T_1 | T_2]$ is a transformation into the union of their range spaces.
Second, principal components returned from running PCA on repeated data samples generally concentrate to the true top principal components for datasets with rapid spectrum drop off.
Work reuse thus proceeds as follows:
DROP maintains a transformation history consisting of the horizontal concatenation of all transformations to this point, computes the SVD of this matrix, and returns the first $k$ columns as the transformation matrix. 

Although this requires an SVD computation, computational overhead is dependent on the size of the history matrix, not the dataset size.
This size is proportional to the original dimensionality $\dvar$ and size of lower dimensional transformations, which are in turn proportional to the data's intrinsic dimensionality and the $TLB$ constraint.
As preserving \emph{all history} can be expensive in practice, 
DROP periodically shrinks the history matrix using DR via PCA. 
We validate the benefit of using work reuse---up to \red{15\%} on real-world data---in \S\ref{sec:experiments}.

\begin{figure*}[t!]
\includegraphics[width=\linewidth]{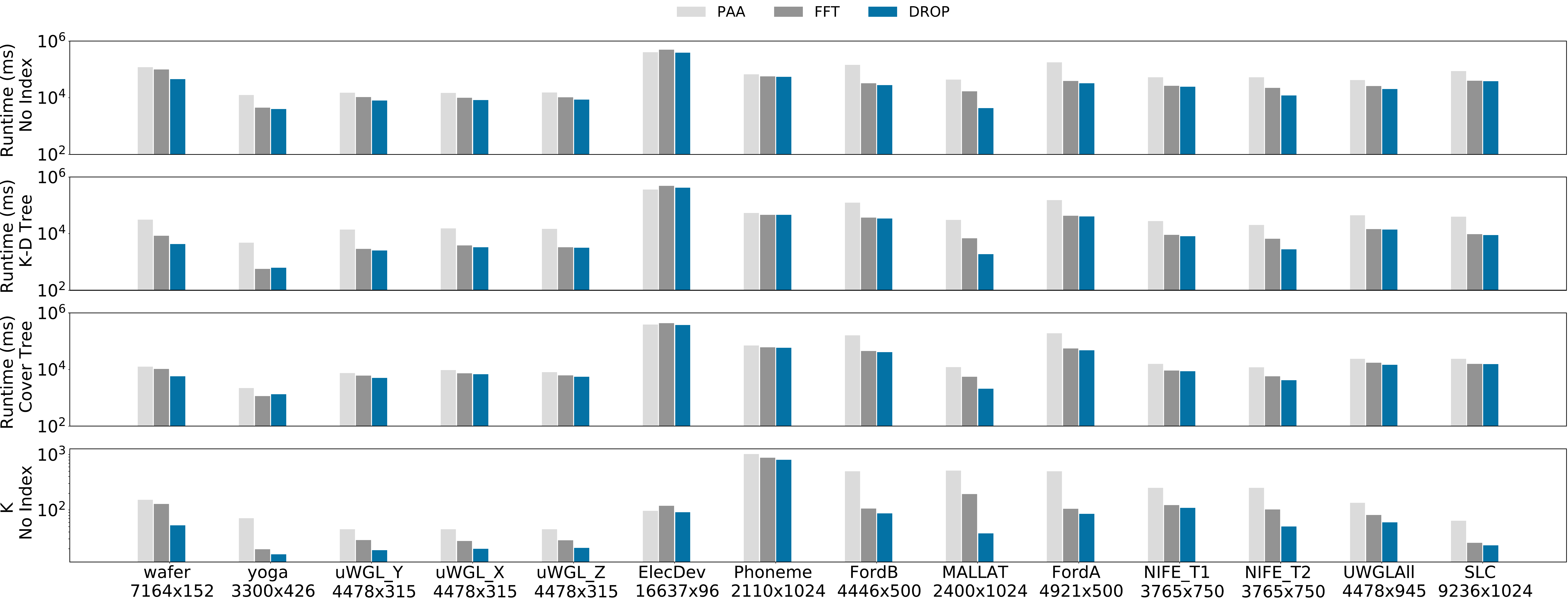}
\caption[]{End-to-End DR and k-NN runtime (top three) and returned lower dimension (bottom) over the largest UCR datasets for three different indexing routines. DROP consistently returns lower dimensional representations than conventional alternatives (FFT, PAA), and is on average faster than PAA and FFT.}
\label{fig:knnAll}
\end{figure*}

\section{Experimental Evaluation}
\label{sec:experiments}

We evaluate DROP's runtime, accuracy, and extensibility. We demonstrate that (1) DROP outperforms PAA and FFT in end-to-end workloads, (2) DROP's optimizations each contribute to performance,  and (3) DROP extends beyond time series.

\subsection{Experimental Setup}
\label{subsec:setup}
\minihead{Implementation} We implement DROP\footnote{\href{https://github.com/stanford-futuredata/DROP}{https://github.com/stanford-futuredata/DROP}} in Java using \red{the multi-threaded Matrix-Toolkits-Java (MTJ) library~\cite{mtj}, and netlib-java~\cite{netlib} linked against Intel MKL~\cite{mkl} for compute-intensive linear algebra operations. 
We use multi-threaded JTransforms~\cite{jtransforms} for FFT, and implement multi-threaded PAA from scratch.}
We use \red{the} Statistical Machine Intelligence and Learning Engine (SMILE) library~\cite{smile} for k-NN {and k-means}. 

\minihead{Datasets} 
We first consider the UCR Time Series Classification Archive~\cite{ucr}, excluding datasets with fewer than 1 million entries, and fewer datapoints than dimensionality, leaving 14 datasets. 
As these are all relatively small time series, we consider four additional datasets to showcase DROP's scalability and generalizability: the MNIST digits dataset~\cite{mnist}, the FMA featurized music dataset~\cite{fma}, a sentiment analysis IMDb dataset~\cite{imdb}, and the fashion MNIST dataset~\cite{fashion}. 

\minihead{DROP Configuration} We use a runtime model for k-NN and k-means computed via polynomial interpolation on data of varying dimension.
While the model is an optional input parameter, any function estimation routine can estimate it given black-box access to the downstream workload.
To evaluate sensitivity to runtime model, we report on the effect of operating without it (i.e., sample until convergence).
We set $TLB$ constraints such that k-NN accuracy remains unchanged, corresponding to $B = 0.99$  for the UCR data.
We use a default sampling schedule that begins with and increases by $1\%$ of the input.
It is possible to optimize (and perhaps overfit) this schedule in future work (\S\ref{subsec:disc}), but we provide a conservative, general schedule as a proof of concept.

\minihead{Baselines} We report runtime, accuracy, and dimensionality compared to FFT, PAA, PCA via SVD-Halko, and PCA via SVD. 
Each computes a transformation over all the data, then performs binary search to identify the lowest dimensionality that satisfies the target $TLB$. 

\minihead{Similarity Search/k-NN Setup} 
We primarily consider k-NN in our evaluation as in~\cite{keogh-study}, but also briefly validate k-means performance.
To evaluate DR performance when used with downstream indexes, we vary k-NN's multidimensional index structure: cover trees~\cite{ctree}, K-D trees~\cite{kdtree}, or no index. 

End-to-end performance depends on the number of queries in the workload, and DROP is optimized for the repeated-query use case. 
Due to the small size of the UCR datasets, we choose a 1:50 ratio of data indexed to number of query points, and vary this index-query ratio in later microbenchmarks and experiments. 
We also provide a cost model for assessing the break-even point that balances the cost of a given DR technique against its indexing benefits.


\subsection{DROP Performance}
\label{subsec:runtime}

We first evaluate DROP's performance compared to PAA and FFT using the time series case study extended from~\cite{keogh-study}. 

\minihead{k-NN Performance} We summarize DROP's results on a 1-Nearest Neighbor classification in Figure~\ref{fig:knnAll}.
We display the end-to-end runtime of DROP, PAA, and FFT for each of the considered index structures: no index, K-D trees, cover trees. 
We display the size of the returned dimension for the no indexing scenario, as the other two scenarios return near \red{identical values.
This occurs as many of the datasets used in this experiment are small and possess low intrinsic dimensionality that DROP quickly identifies
}
We do not display k-NN accuracy as all techniques meet the $TLB$ constraint, and achieve the same accuracy within $1\%$.

On average, DROP returns transformations that are $2.3\times$ and  $1.4\times$ smaller than PAA and FFT, translating to significantly smaller k-NN query time. 
End-to-end runtime with DROP is on average \red{$2.2\times$ and $1.4\times$ (up to $10\times$ and $3.9\times$)} faster than PAA and FFT, respectively, when using brute force linear search,  \red{$2.3\times$ and $1.2\times$ (up to $16\times$ and $3.6\times$)}  faster when using K-D trees, and \red{$1.9\times$ and $1.2\times$ (up to $5.8\times$ and $2.6\times$)} faster when using cover trees.
When evaluating Figure~\ref{fig:knnAll}, it becomes clear that DROP's runtime improvement is data dependent for both smaller datasets, and for datasets that do not possess a low intrinsic dimension (such as Phoneme, elaborated on in \S\ref{subsec:lesion})
Determining if DROP is a good fit for a dataset is an exciting area for future work (\S\ref{subsec:disc}).



\minihead{Varying Index-Query Ratio} DROP is optimized for a low index-query ratio, as in many streaming and/or high-volume data use cases.
If there are many more data points queried than used for constructing an index, but not enough such that expensive, na\"ive PCA is justified, DROP will outperform alternatives. 
A natural question that arises is: at what scale is it beneficial to use DROP?
While domain experts are typically aware of the scale of their workloads, we provide a heuristic to answer this question given rough runtime and cardinality estimates of the downstream task and the alternative DR technique in consideration.

Let $x_d$ and $x_a$ be the per-query runtime of running a downstream task with the output of DROP and a given alternative method, respectively. 
Let $r_d$ and $r_a$ denote the amortized per-datapoint runtime of DROP and the alternative method, respectively. 
Let $n_i$ and $n_q$ the number of indexed and queried points. 
DROP is faster when $n_q x_d + n_i r_d < n_q x_a + n_i r_a$.

\begin{figure}
\includegraphics[width=\linewidth]{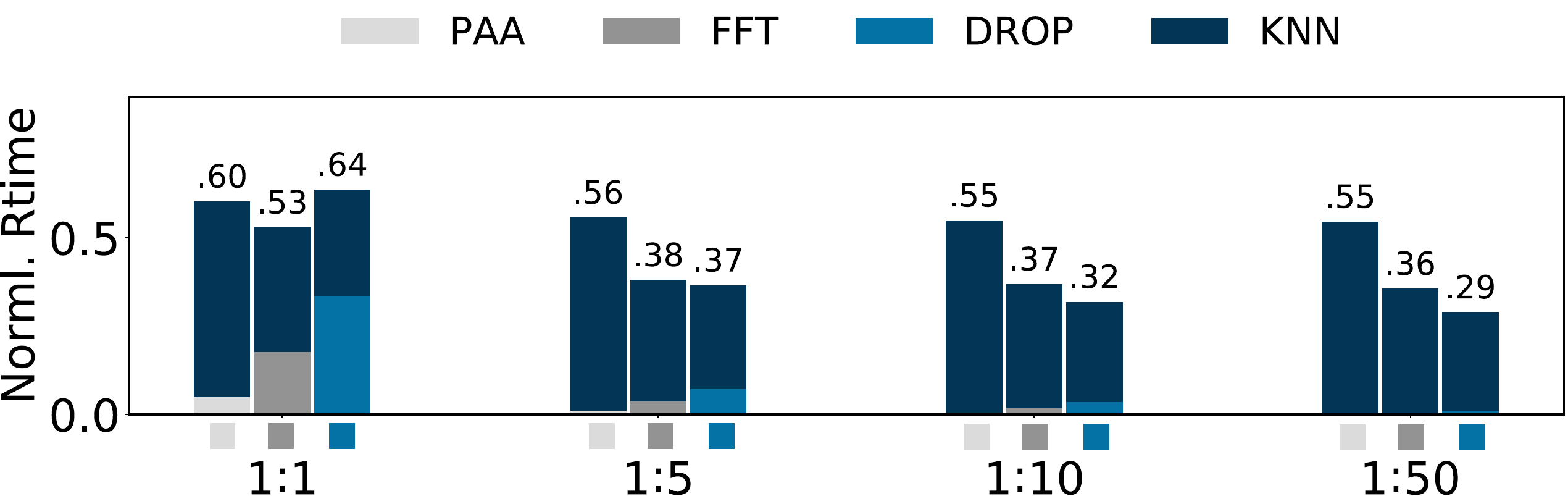}
\caption[]{Effect of decreasing the index-query ratio. As an index is queried more frequently, DROP's relative runtime benefit  increases.}
\label{fig:query}
\end{figure}

To verify, we obtained estimates of the above and empirically validate when running k-NN using cover trees (Figure~\ref{fig:query}).
We first found that in the 1:1 index-query ratio setting, DROP should be slower than PAA and FFT, as observed. 
However, as we decrease the ratio, DROP becomes faster, with a break-even point of slightly lower than 1:3. 
We show that DROP does indeed outperform PAA \red{and FFT} in the 1:5 index-query ratio case, where it is is on average \red{$1.51\times$} faster than PAA and \red{$1.03\times$} faster than FFT. 
As the ratio decreases to 1:50, DROP is up to \red{$1.9\times$} faster than alternatives.

\red{
\minihead{Time Series Similarity Search Extensions}
Given the breadth of research in time series indexing, we evaluate how DROP, a general operator for PCA, compares to time series indexes. 
As a preliminary evaluation, we consider iSAX2+~\cite{isax}, a state-of-the-art indexing tool, in a 1:1 index-query ratio setting, using a publicly available Java implementation~\cite{isaxcode}. 
While these indexing techniques also optimize for the low index-query ratio setting, we find index construction to be a large bottleneck in these workloads. 
For iSax2+, index construction is on average $143\times$ (up to $389\times$) slower than DR via DROP, but is on average only $11.3\times$ faster than k-NN on the reduced space.  However, given high enough query workload, these specialized techniques will surpass DROP.

We also verify that DROP is able to perform well when using downstream similarity search tasks relying on alternative distance metrics, namely, Dynamic Time Warping (DTW)---a commonly used distance measure in the literature~\cite{isaxorig}. 
As proof-of-concept, we implement a 1-NN task using DTW with a 1:1 index-query ratio, and find that even with this high ratio, DROP provides on average $1.2\times$ and $1.3\times$ runtime improvement over PAA and FFT, respectively.

}

\subsection{Ablation Study}
\label{subsec:lesion}

We perform an ablation study of the runtime contributions of each of DROP's components compared to baseline SVD methods. 
We only display the results of k-NN with cover trees; the results hold for the other indexes.
We use a 1:1 index-query ratio \red{with data inflated by 5$\times$} to better highlight the effects of each contribution to the DR routine.

\begin{figure}
\includegraphics[width=\linewidth]{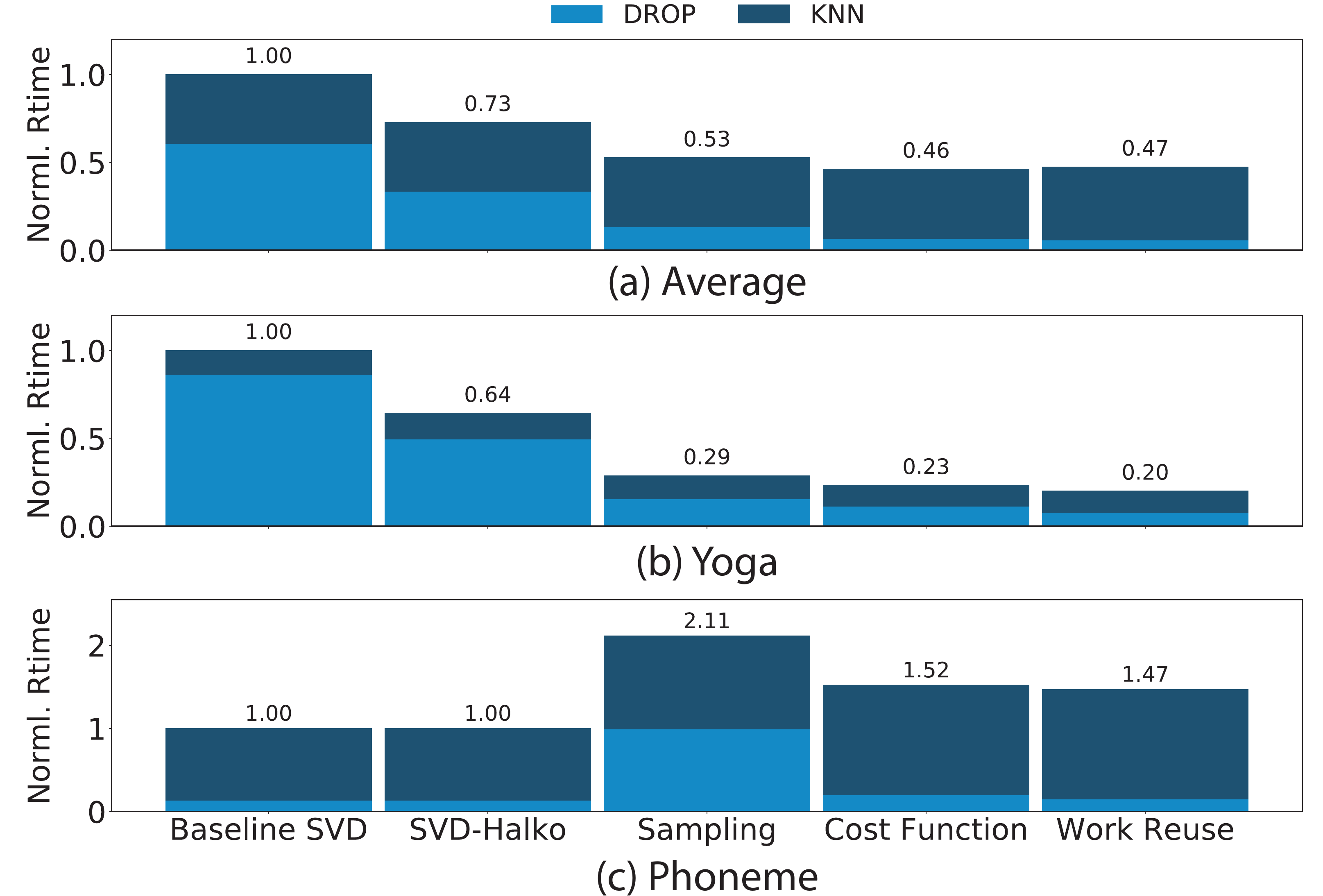}
\caption[]{Ablation Study demonstrating average optimization improvement (a), and sample datasets that are amenable to (b) and operate poorly (c) with DROP}
\label{fig:lesion}
\end{figure}

Figure~\ref{fig:lesion} first demonstrates the boost from using SVD-Halko over a na\"ive implementation of PCA via SVD, which comes from not computing the full transformation a priori, incrementally binary searching as needed. 
It then shows the runtime boost obtained from running on samples until convergence, where DROP samples and terminates after the returned lower dimension from each iteration plateaus.
This represents the na\"ive sampling-until-convergence approach that DROP defaults to sans user-specified cost model.
We finally introduce cost based optimization and work reuse.
Each of these optimizations improves runtime, with the exception of work reuse, which has a negligible impact on average but disproportionately impacts certain datasets. 

Work reuse here typically slightly affects end-to-end runtime as it is useful primarily when a large number of DROP iterations are required.
We also observe this behavior on certain small datasets with moderate intrinsic dimensionality, such as the yoga dataset in Figure~\ref{fig:lesion}b. 
Work reuse provides a $15\%$ improvement over cost based optimization.

DROP's sampling operates on the premise that the dataset has data-point-level redundancy. 
However, datasets without this structure are more difficult to reduce the dimensionality of.
Phoneme is an example of one such dataset (Figure~\ref{fig:lesion}c).  
In this setting, DROP \red{incrementally examines a large proportion of data before enabling cost-based optimization,} resulting in a performance penalty.

\begin{figure}
\includegraphics[width=\linewidth]{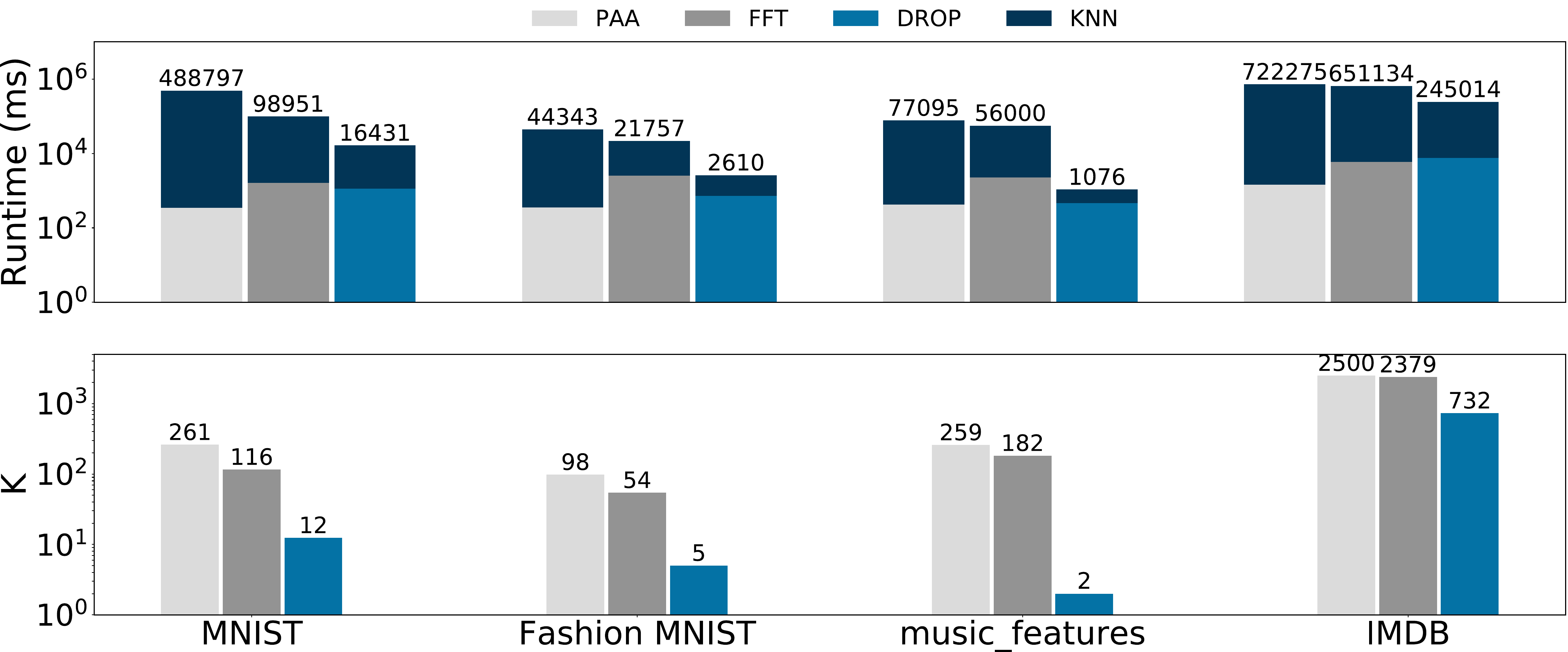}
\caption[]{End-to-End k-NN runtime (top) and returned dimension $k$ (bottom) over four non-time-series datasets spanning text, image, and music }
\label{fig:beyond}
\end{figure}

\red{
\subsection{Beyond Time Series}
\label{subsec:nonts}

We consider generalizability beyond our initial case study along two axes: data domain and downstream workload. 

\subsubsection*{Data Domain}
We examine classification/similarity search workloads across image classification, music analysis, and natural language processing. 
}
We repeat the k-NN retrieval experiments with a 1:1 index-query ratio.
We use the MNIST hand-written digit image dataset of 70,000 images of dimension 784 (obtained by flattening each $28 \times 28$-dimensional image into a single vector~\cite{mnist}, combining both the training and testing datasets); FMA's featurized music dataset, providing 518 features across 106,574 music tracks; a bag-of-words representation of an IMDb sentiment analysis dataset across 25,000 movies with 5000 features~\cite{imdb}; \red{Fashion MNIST's 70,000 images of dimension 784~\cite{fashion}}.  
We present our results in Figure~\ref{fig:beyond}.
As these datasets are larger than those in~\cite{ucr}, DROP's ability to find a $TLB$-preserving low dimensional basis is more valuable as this more directly translates to significant reduction in end-to-end runtime---up to \red{a 7.6 minute wall-clock improvement in MNIST, 42 second improvement in Fashion MNIST, 1.2 minute improvement in music features, and 8 minute improvement in IMDb compared to PAA. 
These runtime effects will only be amplified as the index-query ratio decreases, to be more typical of the repeated-query setting. 
For instance, when we decrease the ratio to 1:5 on the music features dataset, DROP provides a 6.1 and 4.5 minute improvement compared to PAA and FFT, respectively. 
}

\red{
\subsubsection*{Downstream Workload}
To demonstrate the generalizability of  DROP's pipeline as well as black-box runtime cost-model estimation routines, we extend our pipeline to perform a k-means task over the MNIST digits dataset. 
We fit a downstream workload runtime model as we did with k-NN, and operate under a 1:1 index-query ratio. 
DROP terminates in 1488ms, which is 16.5$\times$ and 6.5$\times$ faster than PAA and FFT. 
}

\section{Conclusion and Future Work}
\label{subsec:disc}

DROP provides a first step in bridging the gap between quality and efficiency in DR for downstream \red{analytics}.
However, there are several avenues to explore for future work, such as sophisticated sampling methods and streaming execution:


DROP's efficiency is determined by the dataset's spectrum; MALLAT, with the sharpest drop-off, performs extremely well, and Phoneme, with a near uniform distribution, does not.
Datasets such as Phoneme perform poorly under the default configuration as we enable cost-based optimization after reaching a feasible point.
Thus, DROP spends a disproportionate time sampling (Fig.~\ref{fig:lesion}c). 
Extending DROP to determine if a dataset is amenable to aggressive sampling is an exciting area of future work. 
For instance, recent theoretical results that use sampling to estimate spectrum, even when the number of samples is small in comparison to the input dimensionality~\cite{estspec}, can be run alongside DROP.





In a streaming setting, with a stationary input distribution, users can extract fixed-length sliding windows from the source and apply DROP's transformation over these segments as they arrive. 
Should the data distribution not be stationary, DROP can be retrained in one of two ways. 
First, DROP can make use of the wide body of work in changepoint or feature drift detection~\cite{cp1} to determine when to retrain. 
Alternatively, DROP can maintain a reservoir sample of incoming data~\cite{reservoir}, tuned to the specific application, and retrain if the metric of interest no longer satisfies user-specified constraints. 
Due to DROP's default termination condition, cost-based optimization must be disabled until the metric constraint is achieved to prevent early termination.

\section*{Acknowledgements}
We thank the members of the Stanford InfoLab as well as Aaron Sidford, Mary Wootters, and Moses Charikar for valuable feedback.  
We also thank the creators of the UCR classification archive for their diverse set of time series.
This research was supported in part by affiliate members and other supporters of the Stanford DAWN project---Ant Financial, Facebook, Google, Intel, Microsoft, NEC, SAP, Teradata, and VMware---as well as Toyota Research Institute, Keysight Technologies, Northrop Grumman, Hitachi, and the NSF Graduate Research Fellowship grant DGE-1656518.

\bibliographystyle{ACM-Reference-Format}
{\footnotesize
\bibliography{drop}}
\end{document}